\documentclass[aps,prl,notitlepage,twocolumn,10pt,a4paper]{revtex4-1}

\usepackage{xcolor,graphicx,ulem,soul}
\usepackage{amsmath,amssymb}
\usepackage[colorlinks=true,urlcolor=blue,citecolor=blue,linkcolor=blue]{hyperref}
\usepackage{braket}
\usepackage{cancel}

\newcommand{\cdag}{\hat{c}^{\dagger}}
\newcommand{\dddag}{\hat{d}^{\dagger}}
\newcommand{\xdag}{\hat{x}^{\dagger}}
\newcommand{\ydag}{\hat{y}^{\dagger}}
\newcommand{\adag}{\hat{a}^{\dagger}}
\newcommand{\bdag}{\hat{b}^{\dagger}}
\newcommand{\vac}{\ket{\text{vac}}}
\newcommand{\ud}{\textrm{d}}

\begin{document}


\title{Two-color Bell states heralded via entanglement swapping}


\author{Sofiane Merkouche$^{*1}$, Val\'{e}rian Thiel$^2$, Alex O.C. Davis$^{2,3}$, and Brian J. Smith$^{1,2}$}
\email[]{Corresponding author: sofianem@uoregon.edu}
\affiliation{$^1$ Oregon Center for Optical, Molecular, and Quantum Science, and Department of Physics, University of Oregon, Eugene, OR 97403\\
$^2$ Department of Physics, University of Oxford, Denys Wilkinson Building, Keble Road, Oxford, OX1 3RH, United Kingdom\\
$^3$Laboratoire Kastler-Brossel, UPMC-Sorbonne Universit\'{e}s, CNRS, ENS-PSL  Research University, Coll\`{e}ge de France, 75005 Paris, France}


\date{\today}

\begin{abstract}
We report on an experiment demonstrating entanglement swapping of time-frequency entangled photons. We perform a frequency-resolved Bell-state measurement on the idler photons from two independent entangled photon pairs, which projects the signal photons onto a two-color Bell state. We verify entanglement in this heralded state using two-photon interference and observing quantum beating without the use of filters, indicating the presence of two-color entanglement. Our method could lend itself to use as a highly-tunable source of frequency-bin entangled single photons.
\end{abstract}


\maketitle


\section{Introduction}
Entanglement is one of the most distinguishing features of quantum mechanics, as well as being a crucial resource for quantum information science. An important technique to aid in harnessing this resource is entanglement swapping, which enables entanglement of distant quantum systems and thereby the long-range distribution of quantum correlations, in addition to shedding light on the fundamental nature and extent of quantum non-locality. Entanglement swapping has been experimentally demonstrated using photons entangled in their polarization \cite{Pan1998}, spatial \cite{Zhang2017}, and temporal \cite{Halder2007EntanglingTimemeasurement} degrees of freedom. Frequency entanglement swapping is rather more difficult due to the technical challenges involved in verifying frequency entanglement \cite{Barbieri2017}.

Our frequency entanglement swapping experiment \cite{merkouche2018} is a close analogue to the time-bin experiment in reference \cite{Halder2007EntanglingTimemeasurement}. There, a time-resolved Bell-state measurement is performed on photons from continuously-entangled pairs, heralding a time-bin entangled Bell state. In our experiment, a frequency-resolved Bell-state measurement is performed with the use of narrowband filters, heralding a frequency-bin Bell state. The time-bin experiment, however, uses non-local Franson interference \cite{Franson1989} to verify entanglement in the heralded state, which for our analogue would require a time-varying phase modulation to implement a frequency shear \cite{Olislager2010}. Instead we use two-photon Hong-Ou-Mandel (HOM) interference, whereby observation of photon antibunching corresponds to an anti-symmetric two-photon wavefunction and confirms the presence of entanglement \cite{Fedrizzi2009}.

\section{Theory}
The output state of a single spontaneous parametric down conversion (SPDC) source can be expressed as

\begin{equation}
	\ket{\psi}=\sum_{n=0}^\infty \frac{\sqrt{\eta}^n}{n!}\Big(\int \ud\omega_s \ud\omega_i f(\omega_s, \omega_i) \adag(\omega_s) \bdag(\omega_i)\Big)^n\ket{\mathrm{vac}}
\end{equation}
where $\adag(\omega)\ (\bdag(\omega))$ creates a photon with frequency ${\omega}$ in the signal (idler) mode, and $\eta$ is a parameter associated with the nonlinearity of the interaction and quantifies the probability of generating a photon pair from a pump photon.  The function $f(\omega_s,\omega_i)$ is the normalized complex joint spectral amplitude (JSA), and its modulus squared, $|f(\omega_s,\omega_i)|^2$, is the two-photon probability density for detecting the signal photon at frequency $\omega_s$ and the idler photon at $\omega_i$. The state contains spectral entanglement when the JSA is not factorable; that is, when $f(\omega_s,\omega_i) \neq f_s(\omega_s)f_i(\omega_i)$. We assume that the process is single-mode in the polarization and transverse spatial degrees of freedom, so that only the time-frequency degrees of freedom are relevant.

\begin{figure}[t]
	\centering
	\includegraphics[width=.8\linewidth]{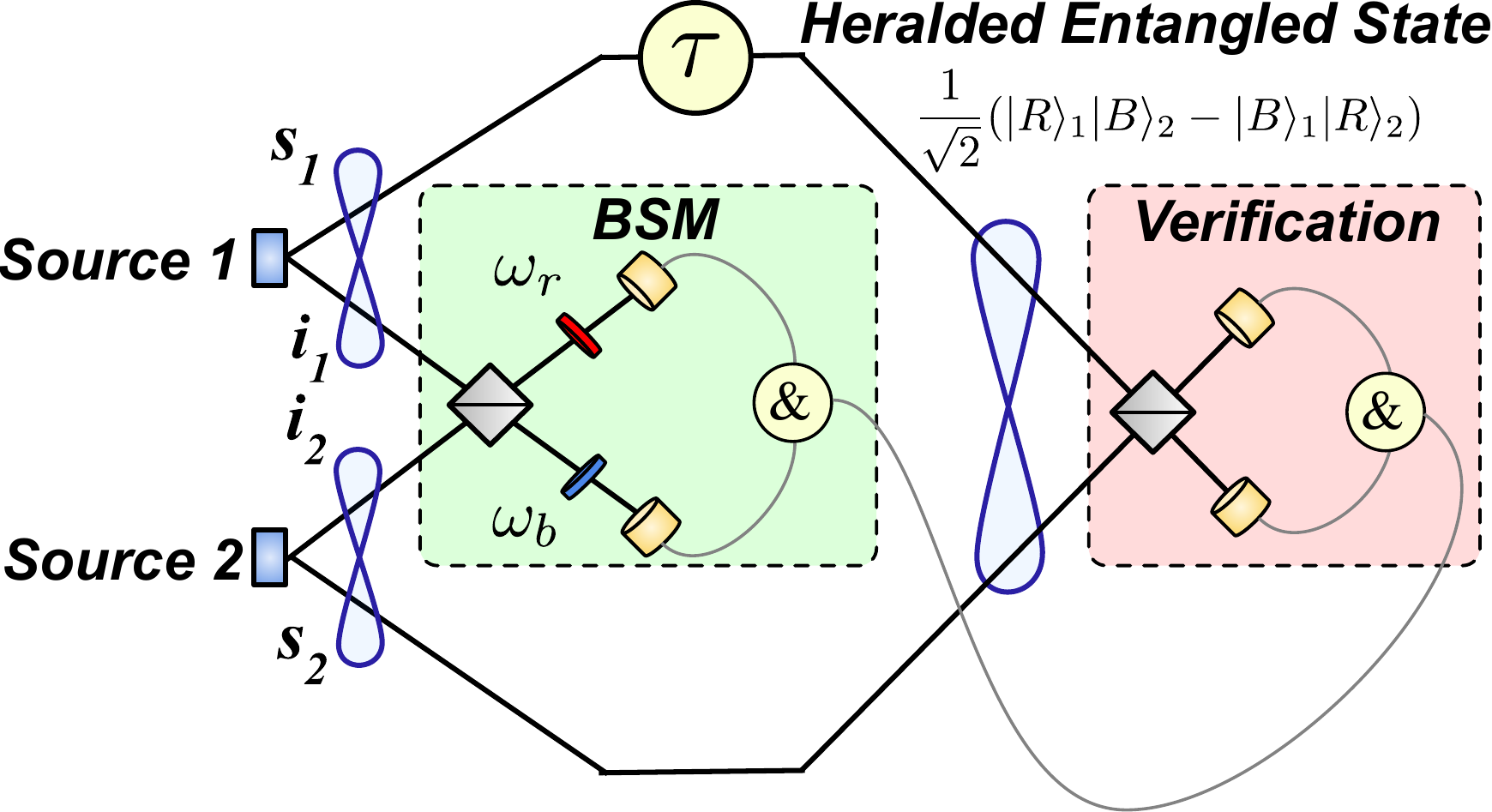}
	\caption{Entanglement swapping scheme. Idler photons from two independent sources of frequency-entangled photon pairs are interfered at a 50-50 beamsplitter, and detected at frequencies $\omega_r$ and $\omega_b$, projecting the signal photons onto the singlet Bell state. Entanglement is verified by two-photon interference.}\label{scheme}
\end{figure}

The experiment we describe is depicted schematically in Fig. \ref{scheme}. We consider two independent but identical SPDC sources, described by a tensor product state $\ket{\psi_1} \otimes \ket{\psi_2}$. Because this is a four-photon experiment, we restrict our attention to the second-order ($\mathcal{O}(\eta)$) contribution in the expansion. If we assume the sources to be mutually incoherent (see Supplemental), this contribution may be written as

\begin{equation}
	\hat{\rho}_\eta = \frac{2}{3}\left(\ket{\psi_{12}}\bra{\psi_{12}} +\frac{1}{4}\ket{\psi_{11}}\bra{\psi_{11}}+\frac{1}{4}\ket{\psi_{22}}\bra{\psi_{22}}\right)
\end{equation}
where
\begin{equation}
	\begin{gathered}
		\ket{\psi_{mn}}=\int \ud\omega_s\ud\omega_i\ud\omega_s'\ud\omega_i'  f(\omega_s,\omega_i)\hat{a}_m^{\dagger}(\omega_s)\hat{b}_m^{\dagger}(\omega_i)\\ \times f(\omega'_s,\omega'_i)\hat{a}_n^{\dagger}(\omega'_s)\hat{b}_n^{\dagger}(\omega'_i)\ket{\mathrm{vac}}
	\end{gathered}
\end{equation}
with $m,\ n \in \{1,2\}$, and where the subscripts 1 and 2 denote the first and second SPDC source, respectively. The first term corresponds to a single pair emission from each source, while the other two terms correspond to a double-pair emission from either one of the sources. 

Entanglement swapping requires performing a partial Bell-state measurement (BSM) on the idler fields $b_1$ and $b_2$, which is achieved by interfering the fields at a 50:50 beamsplitter and performing a frequency-resolved coincident detection at the output, using narrowband filters centered at frequencies $\omega_r$ and $\omega_b$. This measurement projects the input idler fields onto the two-color singlet Bell state $\ket{\psi^-}=\frac{1}{\sqrt{2}} (\ket{\omega_r}_{b1}\ket{\omega_b}_{b2}-\ket{\omega_b}_{b1}\ket{\omega_r}_{b2})$, and the state heralded in the signal fields becomes (see Supplemental)

\begin{equation}
	\begin{gathered}
		\label{state}
		\hat{\rho}_H=\frac{1}{2}\ket{\Psi^-}\bra{\Psi^-}+\frac{1}{4}\ket{\Psi_1}\bra{\Psi_1}+\frac{1}{4}\ket{\Psi_2}\bra{\Psi_2}
	\end{gathered}
\end{equation}
where
\begin{equation}
	\begin{gathered}
		\ket{\Psi^-}=\frac{1}{\sqrt{2}}\left(\ket{R}_1\ket{B}_2 - \ket{B}_1\ket{R}_2\right),\\
		\ket{\Psi_j}=\ket{R}_j\ket{B}_j,
	\end{gathered}
\end{equation}
for $j \in \{1, 2\}$. Here we have defined
\begin{equation}
	\begin{gathered}
		\ket{R}_j = \int \ud\omega \phi_R(\omega) \adag_j(\omega)\vac, \\
		\ket{B}_j = \int \ud\omega \phi_B(\omega) \adag_j(\omega)\vac,
	\end{gathered}
\end{equation}
where $\phi_{R(B)} (\omega) = f( \omega,\omega_{b(r)})$ can be approximated by Gaussian functions as $\phi_{R(B)}(\omega)=\exp{[-(\omega-\omega_{R(B)})^2/2\sigma^2]}$.
The heralded two-color Bell-state is contained in the $\ket{\Psi^-}$ term of (\ref{state}), which arises from the generation of a pair of photons in each source. The other terms, which are due to double-pair emissions, occur in all entanglement swapping experiments to date which rely on SPDC sources and linear optics \cite{Kok2000}. In previous experiments, the double-pair contributions are neglected because the entangled state is post-selected by four-photon coincidences. Because our entanglement verification method relies on two-photon interference of the signal fields, we are required to take these contributions into account.

\begin{figure}[t]
	\centering
	\includegraphics[width=\linewidth]{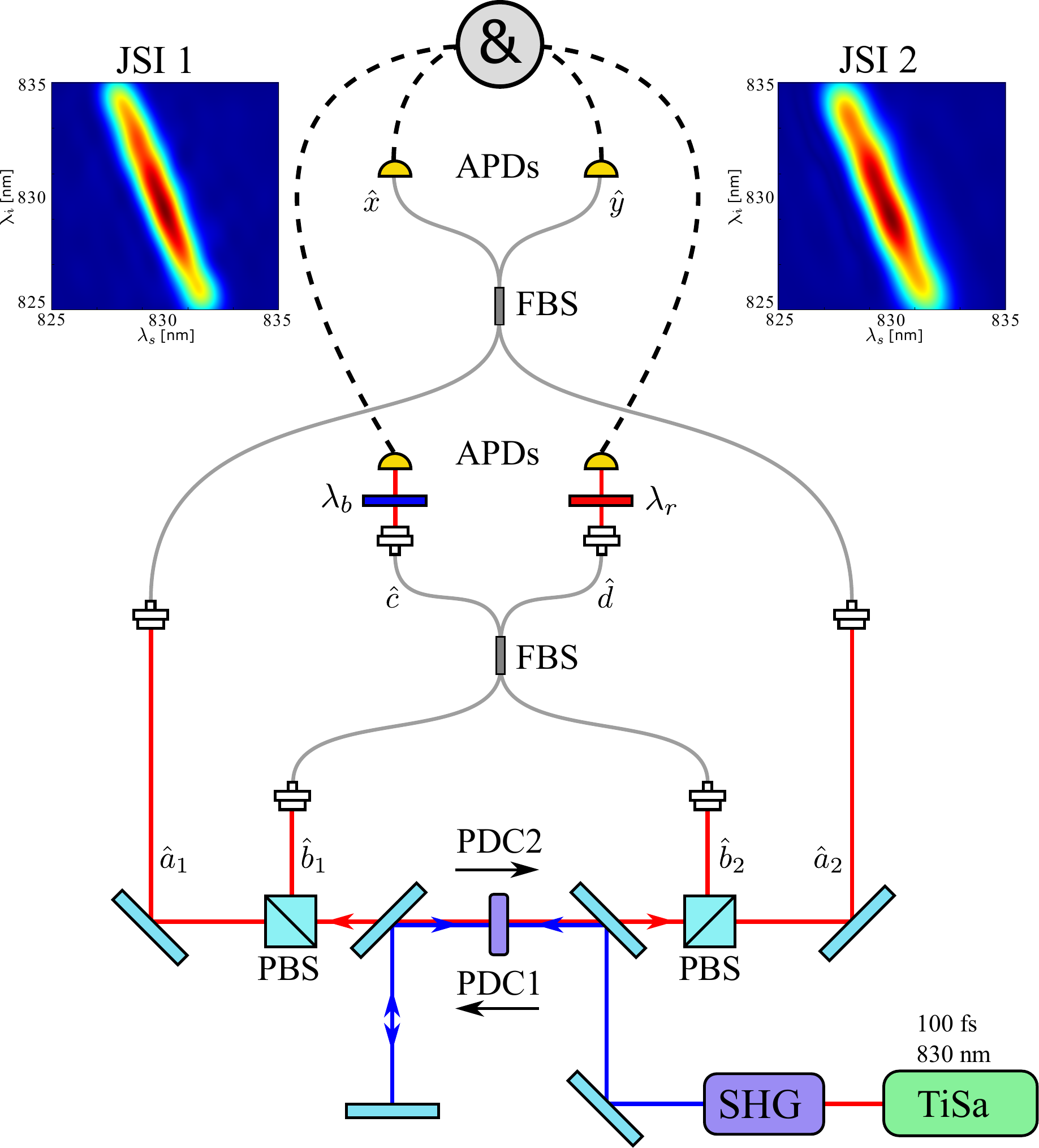}
	\caption{Experimental setup - see main text for description. PBS- polarizing beam splitter; PDC- parametric down conversion; SHG- second harmonic generation; APD: avalanche photodiode; FBS- fiber beamsplitter. Top: joint spectral intensities for sources 1 and 2.} \label{fig:setup}
\end{figure}

In order to show that the $\ket{\Psi^-}$ component of the state is indeed entangled and not merely anticorrelated, two-photon interference is used, whereby the heralded photons are detected in coincidence at the output of a 50:50 beamsplitter, as a function of relative arrival time delay $\tau$. For the input state $\rho_H$, the coincidence probability is given by (see Supplemental)

\begin{equation}
	\label{coinc}
	P_{\text{CC}}(\tau)=\big(\frac{1}{2} + \frac{1}{4}e^{-\tau^2 \sigma^2/2}\cos(\omega_B - \omega_R)\tau\big).
\end{equation}
The oscillations at the frequency difference are due to coherence between the two terms of $\ket{\Psi^-}$. Note that these are observed without filtering in the signal modes. This is similar to other recent experiments producing frequency-bin entangled photons \cite{Ramelow2009} \cite{Jin2018}, with the main difference being that the oscillations there are observed with unit visibility, since only the state $\ket{\Psi^-}$ contributes to coincidences. In our case, the terms due to double pair emission contribute incoherently to a background, reducing the maximum predicted visibility to 1/2. It should be noted that photon antibunching itself, where the coincidence probability exceeds that of the random baseline of 1/2, is indicative of antisymmetry, and thus entanglement, in the two-photon wavefunction \cite{Fedrizzi2009}.

\section{Experiment} 
The light source for the experiment consists of ultrashort (100 fs) pulses from a titanium-doped sapphire (Ti:Sapph) laser oscillator at a central wavelength of 830 nm and a repetition rate of 80 MHz. These pulses are frequency-doubled in a 1-mm long birefringent $\mathrm{BiB_3O_6}$ crystal (BiBO) to generate the blue (415 nm) pump for the parametric downconversion (PDC) sources. PDC occurs at a second, 2.5-mm long BiBO, which is double-passed to probabilistically create a pair of frequency-entangled photons on the first pass (source 1), and on the second pass (source 2). This double-pass configuration ensures that the two sources are identical. Type II phase matching permits the deterministic separation of the signal and idler fields using polarizing beamsplitters (PBS), after the blue pump has been filtered out using dichroic mirrors. Signal and idler photons from both sources are collected into polarization-maintaining single-mode fibers (PM fibers) and directed to the remainder of the set-up for analysis and entanglement swapping. We measure a pair detection rate of 100 kHz from each source, uncorrected for losses.

\begin{figure}	 
	\centering
	\includegraphics[width=\linewidth]{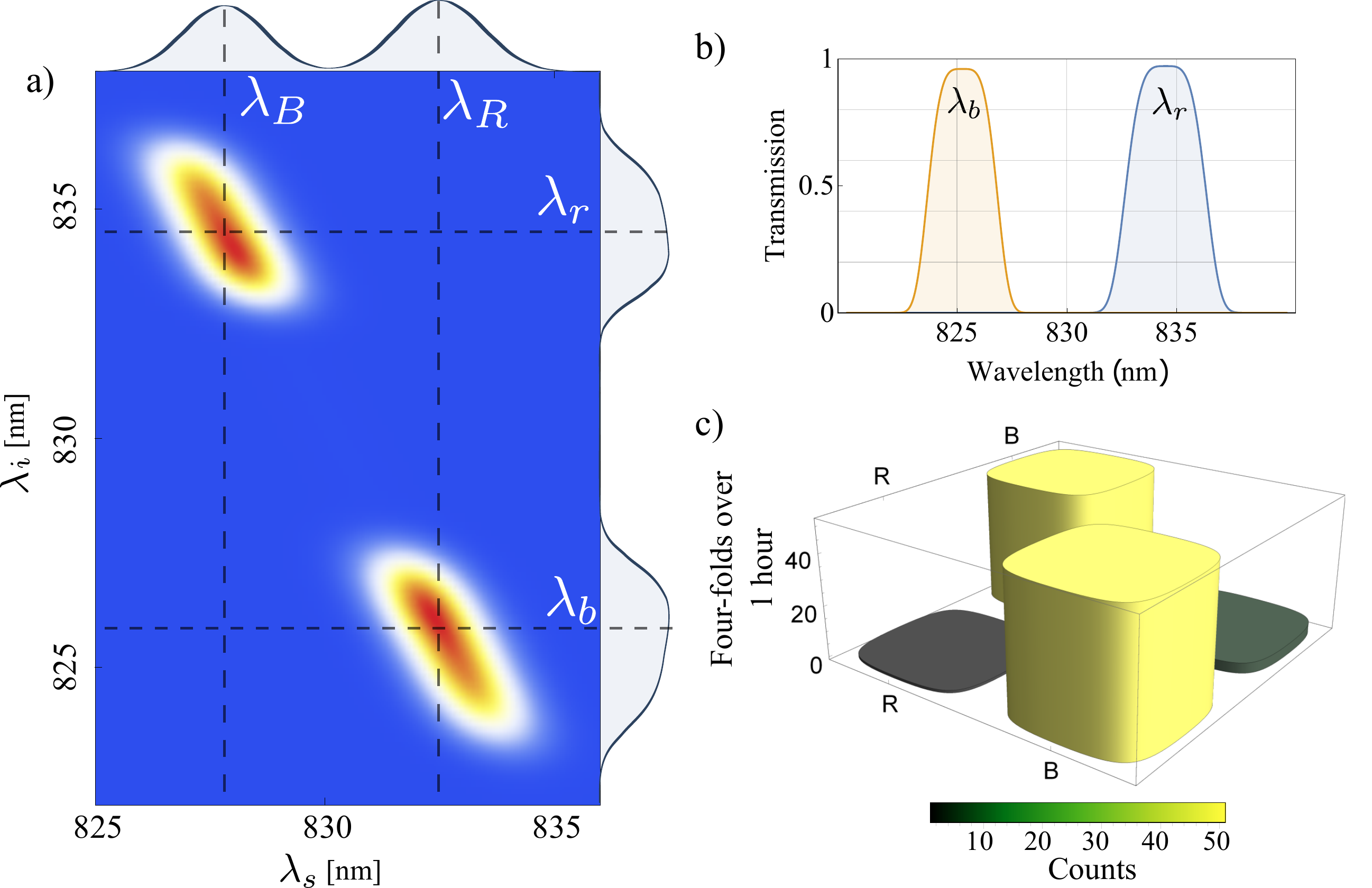}
	\caption{a) Simulated filtered joint spectral intensity using experimental data. b) Measured transfer functions of the filters after the idler FBS. c) Experimentally measured heralded signal state, which resembles a).}\label{fig:herald}
\end{figure}

The joint spectral intensities of each source are measured efficiently using a time-of-flight spectrometer \cite{Davis2016}, consisting of two highly dispersive chirped fiber Bragg gratings (CFBG) followed by two low-timing-jitter single photon detectors. The CFBG imparts a frequency-dependent delay on the incoming photons, thus mapping frequency onto time. Time-resolved coincidence detections at the output then provide a direct measure of the joint spectral intensity (see Fig. \ref{fig:setup}, top), which show frequency anticorrelation. The JSI shown in Fig. \ref{fig:setup} are cropped because of the reflectivity bandwidth of the CFBG, but they extend further, since the bandwidth of the marginal spectra were measured at $9.0$ nm and $16.4$ nm full width at half maximum (FWHM) for the signal and idler photon, respectively, using a single-photon-sensitive spectrometer. Assuming negligible phase correlations \cite{davis2018}, we estimate the amount of entanglement in the state by taking the square root of the JSI and calculating the Schmidt number \cite{Law2000}, for which we obtain a value of $K \sim 5$.
     
     
The entangling partial BSM is performed on the idler photons by routing them to a polarization maintaining fiber beamsplitter (FBS) to ensure mode-matching and indistinguishability at the output, while temporal matching is achieved using a free-space delay line. The delays between the two sources were matched by monitoring the arrival time of the single photon packets with the fast APD, and matching them with delay lines placed in the both signal and idler paths of source 1. Sub-picosecond delay matching was subsequently achieved by monitoring the coincidences between the uncorrelated signals and idlers from both sources, at a low coincidence rate ($\sim 4$ kHz), until HOM interference is observed.

\begin{figure}[t]
	\centering
	\includegraphics[width=\linewidth]{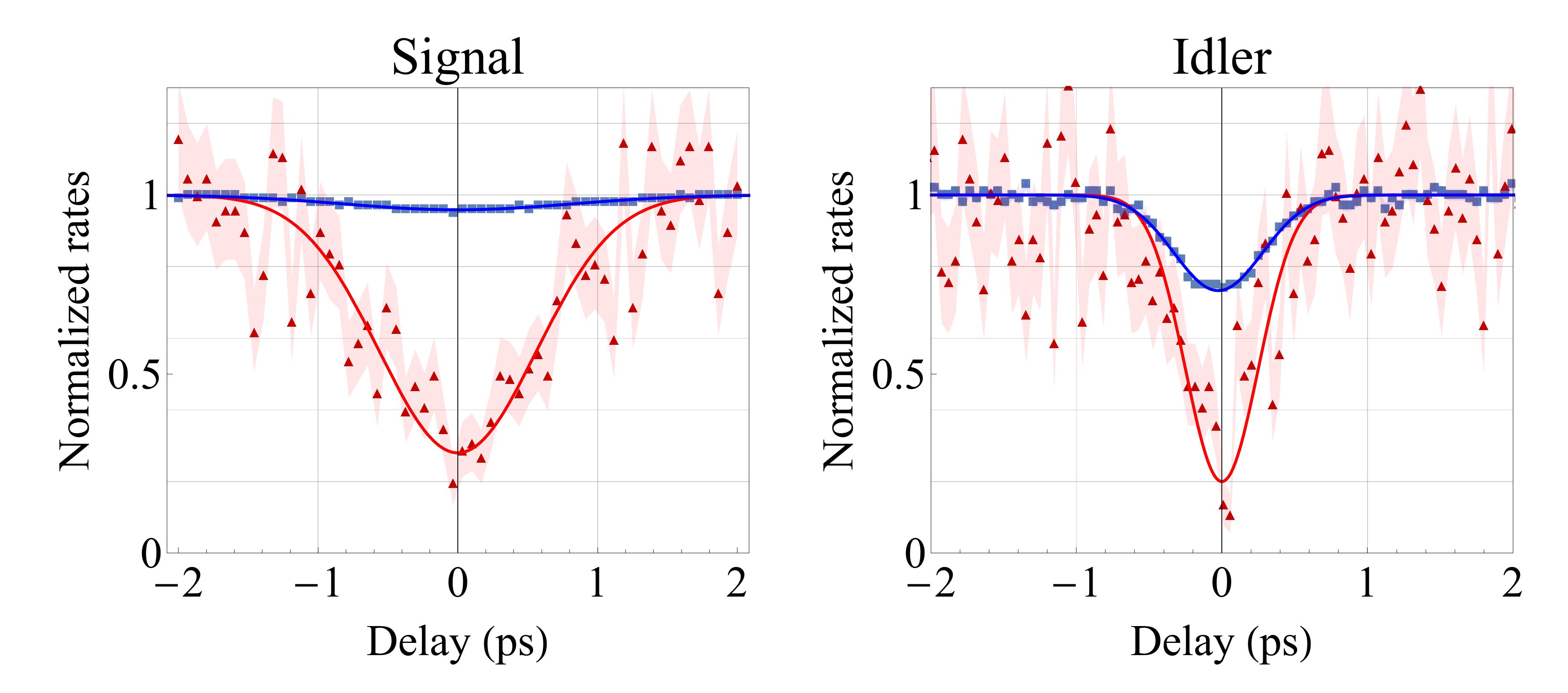}
	\caption{Interferences between uncorrelated signal and idler photons. Square (blue) unheralded coincidences; Triangle (red) heralded coincidences where the other arm is spectrally filtered.}\label{fig:homdips}
\end{figure}

The visibility of the HOM dip was low ($< 25\%$) because the interfering photons are in mixed states due to the spectral entanglement with their respective twin photons. Since filtering the heralds increases the purity of the heralded photons, and since filters are used in the entanglement swapping configuration, we were able to obtain a lower bound on the purity of the heralded photons in the following manner. We angle-tuned our narrowband heralding filters (3 nm FWHM) to the same wavelength (830 nm), and measured the HOM dip visibility, and thus lower bound of the heralded purity, of the signal (idler) photons, when the idler (signal) photons are filtered. This measurement obtains a HOM interference visibility of 72\% for the signal photons, and 80\% for the idler photons, both of which are plotted in Fig. \ref{fig:homdips}. For the entanglement swapping, the idler photons are detected in coincidence using SPCMs at the output of the FBS and with the narrowband filters angle-tuned to $\lambda_r=834.5$ nm and $\lambda_b=825$ nm. This measurement heralds the state $\hat{\rho}_H$ in the signal modes, with $\lambda_R=826.8\ \text{nm}$, $\sigma_\textrm{R}=2.2$ nm FWHM and $\lambda_B = 832.5\ \text{nm}$, $\sigma_\textrm{B}=2.3$ nm FWHM.

To characterize the heralded state, we measure the populations in the $\ket{\Psi^-}$ subspace of the state by spectrally filtered coincidences on the separated signal fields, conditioned on the partial BSM. This measurement method was preferable to the time-of-flight spectrometer due to the prohibitive losses of the latter, combined with the low rate ($\sim$ 1 Hz) of fourfold coincidences. The results are plotted in Fig. \ref{fig:herald}, where frequency anticorrelations can be observed. 

Finally, to verify that the heralded state is not only anticorrelated but entangled, we use two-photon interference by overlapping the signal modes at a 50:50 fiber beamsplitter and observe coincidences at the output while varying the relative arrival time $\tau$ at the input, conditioned on the partial BSM. Observation of modulation of the coincidence rate with a period of $2\pi/(\omega_B-\omega_R)$ as a function of $\tau$, as in expression (\ref{coinc}), is indicative of coherence between the two terms in the $\ket{\Psi^-}$ subspace, and thus of frequency entanglement.

The two-photon interference measurements, collected over a period of 15 hours, are shown in Fig. \ref{fig:fringes}, where the predicted oscillations can be observed. A fit of the interferogram to the function $P(\tau)=A\big(1 + V e^{-\beta\tau^2}\cos\Delta\omega\tau\big)$ gives a period of $540\pm 30$ fs and with an envelope of $1.0 \pm 0.2$ ps, consistent with the measured values of $\omega_{R/B}$ and $\sigma_{R/B}$. The observed visibility of the oscillations V=$0.27 \pm 0.04$ is significantly less than the prediction of 0.5. This can be accounted for mostly by source mismatch, time delay drift, and beamsplitter ratio offset. However, it is sufficient to note that the central interference peak reveals anti-symmetry in the two-photon state, and hence entanglement \cite{Fedrizzi2009}. A second fit is also shown in Fig. \ref{fig:fringes}, which accounts for a relative phase in the heralded state, i.e. $\ket{\Psi^-}  = \frac{1}{\sqrt{2}}(\ket{R}_1\ket{B}_2 + e^{i\phi}\ket{B}_1\ket{R}_2)$, due to a non-zero delay for the herald photons in the BSM. This manifests itself as a phase offset in the interferogram which corresponds to a temporal delay of merely $\tau_i=70$ fs and reducing fringes visibility by an additional 5\%

\begin{figure}
\centering
\includegraphics[width=.85\linewidth]{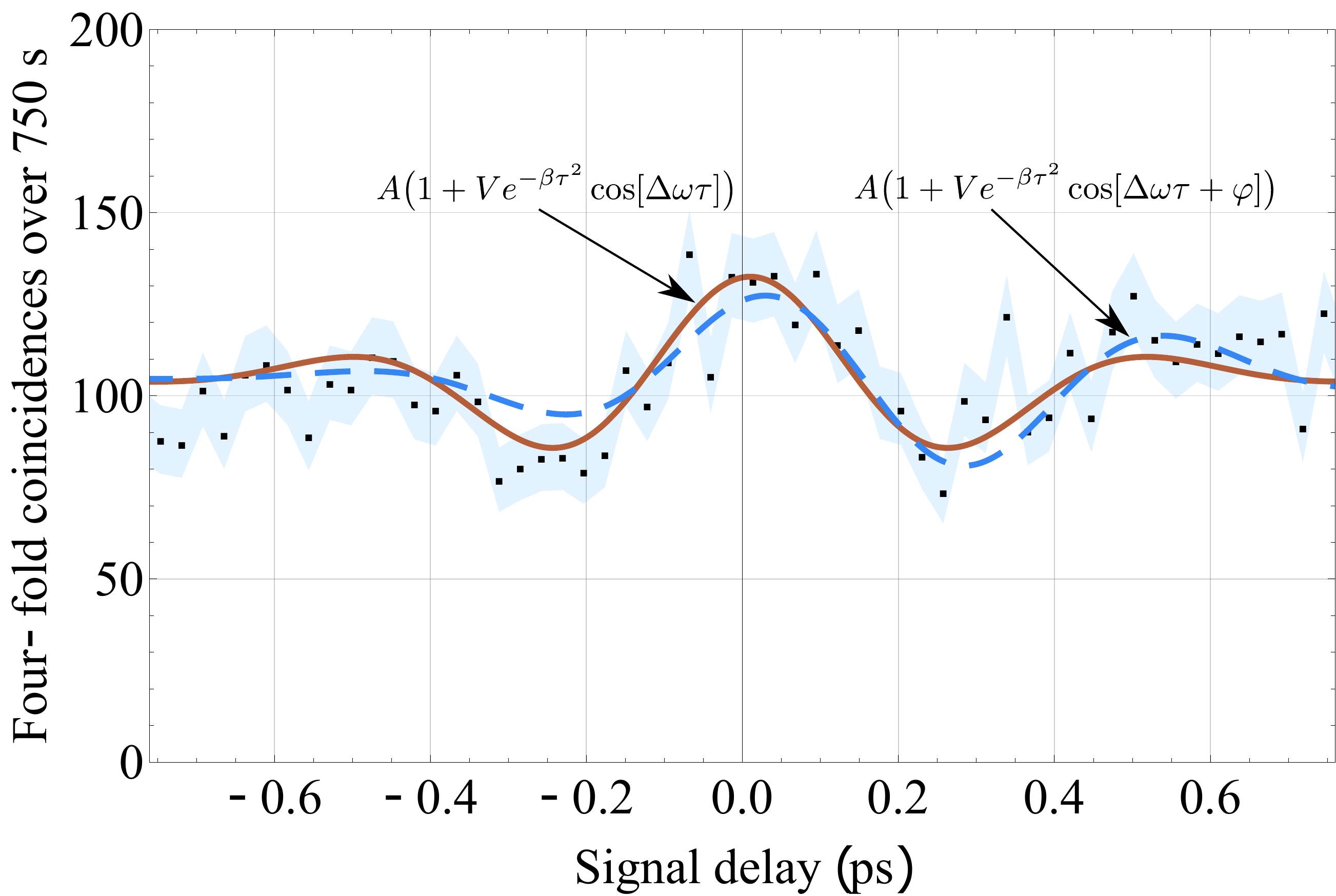}
\caption{Two-photon interference fringes as a function of relative arrival time delay $\tau$ of the heralded signal photons at the beamsplitter, as described in Fig. \ref{scheme}.}\label{fig:fringes} 
\end{figure}

\section{Discussion}
The observed two-photon interference is an unambiguous signature of the presence of frequency-bin entanglement \cite{Ramelow2009}. More precisely, the antibunching itself, where the coincidence probability exceeds 1/2, is a clear indication of anti-symmetric entanglement \cite{Fedrizzi2009}. As is the case in the time-bin experiment in reference \cite{Halder2007EntanglingTimemeasurement} and similar experiments \cite{Pan1998, Zhang2017}, our method is strictly \textit{post-selected} entanglement swapping, where one can only claim an entangled state in the signal modes when separated detectors click in these modes. As has been noted in reference \cite{Kok2000}, it is not correct to interpret the heralded state in equation (\ref{state}) as a preparation of $\ket{\Psi^-}$ with probability $1/2$, as this interpretation commits the partition fallacy: the choice of basis in which to write the density matrix $\hat{\rho}_H$ is arbitrary and does not depend on the underlying physical processes. This is not a limitation for using this technique when the heralded photons are detected non-locally, because only the entangled component of the state contributes to such detections. On the other hand, generation of a pure heralded entangled photon pair from SPDC sources and linear optics is possible with the use of four herald photons \cite{Wagenknecht2010}, or alternately by using a nonlinear interaction such as sum-frequency generation to jointly measure the two herald photons \cite{Sangouard2011FaithfulGeneration} \cite{Vitullo2018}. At the time of writing, a similar experiment has been reported \cite{Graffitti2019} where time-frequency Bell states are generated at the SPDC sources and swapped and verified using the same method we use.


\section{acknowledgments}
	This project has received funding from the European Union's Horizon 2020 research and innovation programme under Grant Agreement No. 665148, the United Kingdom Defense Science and Technology Laboratory (DSTL) under contract No. DSTLX-100092545, and the National Science Foundation under Grant No. 1620822.

\bibliography{biblio}

\onecolumngrid
\pagebreak

\setcounter{equation}{0}

\section{Experimental frequency entanglement swapping: Supplemental materials}

\subsection{Deriving the heralded state}

To derive the state $\hat{\rho}_H$ heralded by the BSM, we begin by writing the SPDC state due to two independent and identical sources as a tensor product

\begin{equation}
\begin{gathered}
\left\{\hat{1} + \sqrt{\eta} \int \ud\omega_s \ud\omega_i f(\omega_s,\omega_i)\adag_1(\omega_s)\bdag_1(\omega_i) + \frac{\eta}{2}\left(\int \ud\omega_s \ud\omega_i f(\omega_s,\omega_i)\adag_1(\omega_s)\bdag_1(\omega_i)\right)^2 + \dots \right\}\\
\otimes\left\{\hat{1} + \sqrt{\eta} \int \ud\omega_s \ud\omega_i f(\omega_s,\omega_i)\adag_2(\omega_s)\bdag_2(\omega_i) + \frac{\eta}{2}\left(\int \ud\omega_s \ud\omega_i f(\omega_s,\omega_i)\adag_2(\omega_s)\bdag_2(\omega_i)\right)^2 + \dots \right\}\vac.
\end{gathered}
\end{equation}
We expand this and keep only terms of order $\eta$, which are responsible for the four-photon contribution:

\begin{equation}
\begin{gathered}
\ket{\Psi_{\mathrm{\eta}}}=\int \ud\omega_s \ud\omega_i \ud\omega_s' \ud\omega_i' f(\omega_s,\omega_i)f(\omega_s',\omega_i') \adag_1(\omega_s) \bdag_1(\omega_i)  \adag_2(\omega_s') \bdag_2(\omega_i')\vac\\
+\frac{1}{2}\int \ud\omega_s \ud\omega_i \ud\omega_s' \ud\omega_i' f(\omega_s,\omega_i)f(\omega_s',\omega_i') \adag_1(\omega_s) \bdag_1(\omega_i)  \adag_1(\omega_s') \bdag_1(\omega_i')\vac\\
+\frac{1}{2}\int \ud\omega_s \ud\omega_i \ud\omega_s' \ud\omega_i' f(\omega_s,\omega_i)f(\omega_s',\omega_i') \adag_2(\omega_s) \bdag_2(\omega_i)  \adag_2(\omega_s') \bdag_2(\omega_i')\vac.
\end{gathered}
\end{equation}
For convenience, we will denote these three terms $\ket{\Psi_{12}}$, $\ket{\Psi_{11}}$, and $\ket{\Psi_{22}}$, so that we have, with the proper renormalization

\begin{equation}
\ket{\Psi_\mathrm{\eta}} = \sqrt{\frac{2}{3}}\left(\ket{\Psi_{12}}+\frac{1}{2}\ket{\Psi_{11}} + \frac{1}{2}\ket{\Psi_{22}}\right), \label{eq:psip}
\end{equation}
and the density matrix for this state is

\begin{equation}
\hat{\rho}_\mathrm{\eta} = \frac{2}{3}\left(\ket{\Psi_{12}}\bra{\Psi_{12}} + \frac{1}{4}\ket{\Psi_{11}}\bra{\Psi_{11}} + \frac{1}{4}\ket{\Psi_{22}}\bra{\Psi_{22}}\right) +\cancel{ \mathrm{cross\ terms}}.
\label{rhop}
\end{equation}
The cross terms correspond to coherence between the terms in $\ket{\Psi_\eta}$, which is ultimately due to the optical phase of the pump. Because our sources are pumped by the same laser, we do indeed expect them to be mutually coherent. However, over the course of a measurement run (several hours), the phase drifts significantly, so it is reasonable to average over it, and thus these cross terms vanish.

A Bell state measurement is performed on the idler photons by interfering them at a beamsplitter and detecting coincidences at the output. The POVM element for this detection is given by

\begin{equation}
\hat{\Pi}_\mathrm{BSM} = \int \ud\omega \ud\omega' |t_r(\omega)|^2 |t_b(\omega')|^2 \cdag(\omega)\dddag(\omega')\vac\bra{\mathrm{vac}}\hat{c}(\omega)\hat{d}(\omega')
\end{equation}
where $t_{r(b)}(\omega)$ is a filter transmission amplitude centered at $\omega_{r(b)}$, and

\begin{equation}
\cdag(\omega) = \frac{\bdag_1(\omega) + \bdag_2(\omega)}{\sqrt{2}},\quad \dddag(\omega') = \frac{\bdag_1(\omega') - \bdag_2(\omega')}{\sqrt{2}}
\end{equation}
are the operators at the output of the beamsplitter. From this, we obtain the heralded state by tracing over the idler modes

\begin{equation}
\hat{\rho}_H = \mathrm{Tr}_b(\hat{\rho}_\eta\hat{\Pi}_\mathrm{BSM}) = \int \ud\omega \ud\omega' \bra{\mathrm{vac}}\hat{b}_1(\omega)\hat{b}_2(\omega') \hat{\rho}_\mathrm{p}\hat{\Pi}_\mathrm{BSM}\bdag_1(\omega)\bdag_2(\omega')\vac \label{eq:trace}
\end{equation}
We take the limit $t_{r(b)}(\omega) \rightarrow \delta(\omega-\omega_{r(b)})$, corresponding to a frequency-resolved measurement of the idler photons. Upon taking the trace, we define the functions

\begin{equation}
\phi_{R}(\omega) = \int \ud\omega' \delta(\omega-\omega_{b}) f(\omega,\omega'),\quad 
\phi_{B}(\omega) = \int \ud\omega' \delta(\omega-\omega_r) f(\omega,\omega').
\end{equation}
The frequency-resolved limit in the measurement of the idler photons implies that the signal photons are heralded in pure states described by the spectral amplitudes $\phi_{R(B)}(\omega)$. We take these to be Gaussian functions of the same width $\sigma$:

\begin{equation}
\phi_{R(B)}(\omega) = \frac{1}{\sqrt{\sigma\sqrt{\pi}}}\exp[-(\omega-\omega_{R(B)})^2/2\sigma^2]
\end{equation}

Taking all this into account, we evaluate the trace term by term in equation \ref{rhop}:

\begin{equation}
\begin{gathered}
\mathrm{Tr}_b(\ket{\Psi_{12}}\bra{\Psi_{12}}\hat{\Pi}_\mathrm{BSM}) = \frac{1}{2}\ket{\Psi^-}\bra{\Psi^-};\\
\ket{\Psi^-} = \frac{1}{\sqrt{2}}\int \ud\omega_1 \ud\omega_2 \big(\phi_R(\omega_1)\phi_B(\omega_2) - \phi_R(\omega_2)\phi_B(\omega_1)\big)\adag_1(\omega_1)\adag_2(\omega_2)\vac,
\end{gathered}
\end{equation}
which gives the swapped entangled state, and

\begin{equation}
\begin{gathered}
\mathrm{Tr}_b(\ket{\Psi_{11(22)}}\bra{\Psi_{11(22)}}\hat{\Pi}_\mathrm{BSM}) = \ket{\Psi_{1(2)}}\bra{\Psi_{1(2)}};\\
\ket{\Psi_{1(2)}} = \int \ud\omega \ud\omega' \phi_R(\omega)\phi_B(\omega')\adag_{1(2)}(\omega)\adag_{1(2)}(\omega')\vac
\end{gathered},
\end{equation}
which correspond to a pair of photons in each arm, one red and one blue. Combining these and renormalizing, we arrive at the heralded state

\begin{equation}
\hat{\rho}_{H} = \frac{1}{2}\ket{\Psi^-}\bra{\Psi^-} + \frac{1}{4}\ket{\Psi_1}\bra{\Psi_1} + \frac{1}{4}\ket{\Psi_2}\bra{\Psi_2}.
\end{equation}

\subsection{Calculating the coincidence probability}

Given the heralded state $\hat{\rho}_H$ of the signal photons, conditioned on the Bell-state measurement on the idler photons, we can calculate the coincidence probability after the second beamsplitter. The measurement operator for an unresolving coincidence detection at the output of the beamsplitter is given by

\begin{equation}
\hat{\Pi} = \int \ud\omega \ud\omega' \xdag(\omega)\ydag(\omega')\vac\bra{\mathrm{vac}}\hat{x}(\omega)\hat{y}(\omega'),
\end{equation}
with

\begin{equation}
\xdag(\omega) = \frac{\adag_1(\omega) + e^{i\omega\tau}\adag_2(\omega)}{\sqrt{2}},\quad \ydag(\omega') = \frac{\adag_1(\omega') - e^{i\omega'\tau}\adag_2(\omega')}{\sqrt{2}},
\end{equation}
where a relative phase is obtained from a relative time delay $\tau$.

We recall that the single-photon wave packets are Gaussian functions normalized such that $\int \ud\omega |\phi_{R(B)}(\omega)|^2 = 1$. The coincidence probability is then given by

\begin{align}
P_{cc}(\tau) &= \mathrm{Tr}(\hat{\rho}_H\hat{\Pi}) = \int \ud\omega \ud\omega' \bra{\mathrm{vac}} \hat{a}_1(\omega)\hat{a}_2(\omega')\ \hat{\rho}_H\hat{\Pi}\ \adag_1(\omega)\adag_2(\omega')\vac \nonumber \\
&= \frac{1}{4} + \frac{1}{4}e^{-\tau^2\sigma^2/2}\cos(\omega_B - \omega_R)\tau + \frac{1}{8}+\frac{1}{8} \nonumber \\
&=\ \frac{1}{2} + \frac{1}{4}e^{-\tau^2\sigma^2/2}\cos(\omega_B - \omega_R)\tau.
\end{align}

\subsection{Full model}

The previous derivation assumes that the inteference filters at the output of the idler FBS are delta functions that herald a pure state in the signal arm. In reality, these filters have a finite bandwidths, and therefore a more complete model should take into account the overlap between the modes at both beamsplitters. In this section we show the derivation that leads to a full description of the process. We also consider that both sources might have a different joint spectrum, leading to decrease in visibility.

In that model, equation \eqref{eq:psip} still holds, but the individual terms are defined as:

\begin{align}
\ket{\Psi_{nm}}=\int \ud\omega_s \ud\omega_i \ud\omega_s' \ud\omega_i' f_n(\omega_s,\omega_i)f_m(\omega_s',\omega_i') \adag_1(\omega_s) \bdag_1(\omega_i)  \adag_2(\omega_s') \bdag_2(\omega_i')\vac,
\end{align}
where we keep track of the source joint spectral amplitude. Computing the heralded state using Eq. \eqref{eq:trace} now takes into account a certain filter transfer function as opposed to projecting in a single frequency bin. The filtered joint spectrum for source $n$ (where $n=1,2$) consists of two individual distributions $\phi_{A,n}(\omega_s,\omega_i)$ and $\phi_{B,n}(\omega_s,\omega_i)$ which contain a spectral phase (mostly linear phase, which is temporal delay). Following the development to the end under these considerations, we then obtain the coincidence probability:

\begin{align}
P_{cc} = \frac{1}{2}+\frac{1}{4} \textrm{Re}\Big\{ \int \ud^2\omega\ \phi_{A,1}^\ast(\omega_s,\omega_i) \phi_{A,2}(\omega_s,\omega_i) \int \ud^2\omega\ \phi_{B,1}(\omega_s,\omega_i) \phi_{B,2}^\ast(\omega_s,\omega_i) \Big\}
\end{align}
which clearly shows that the interference pattern is defined by overlap integrals between the two individual spectral distributions from each source. By identifying these distributions as centered around their respective center frequencies on both signal and idler and considering some delay for each path of the interferometer, it is possible to show that the overlap integrals correspond to cross-correlations in time. This allows to model how mismatch between the two sources are impacting the measurement. Ultimately, the most important parameter that affects visibility of the fringes is found to be the delay between the two idler photons that are used to herald the measurement.

\begin{figure}
	\centering
	\includegraphics[width=.9\linewidth]{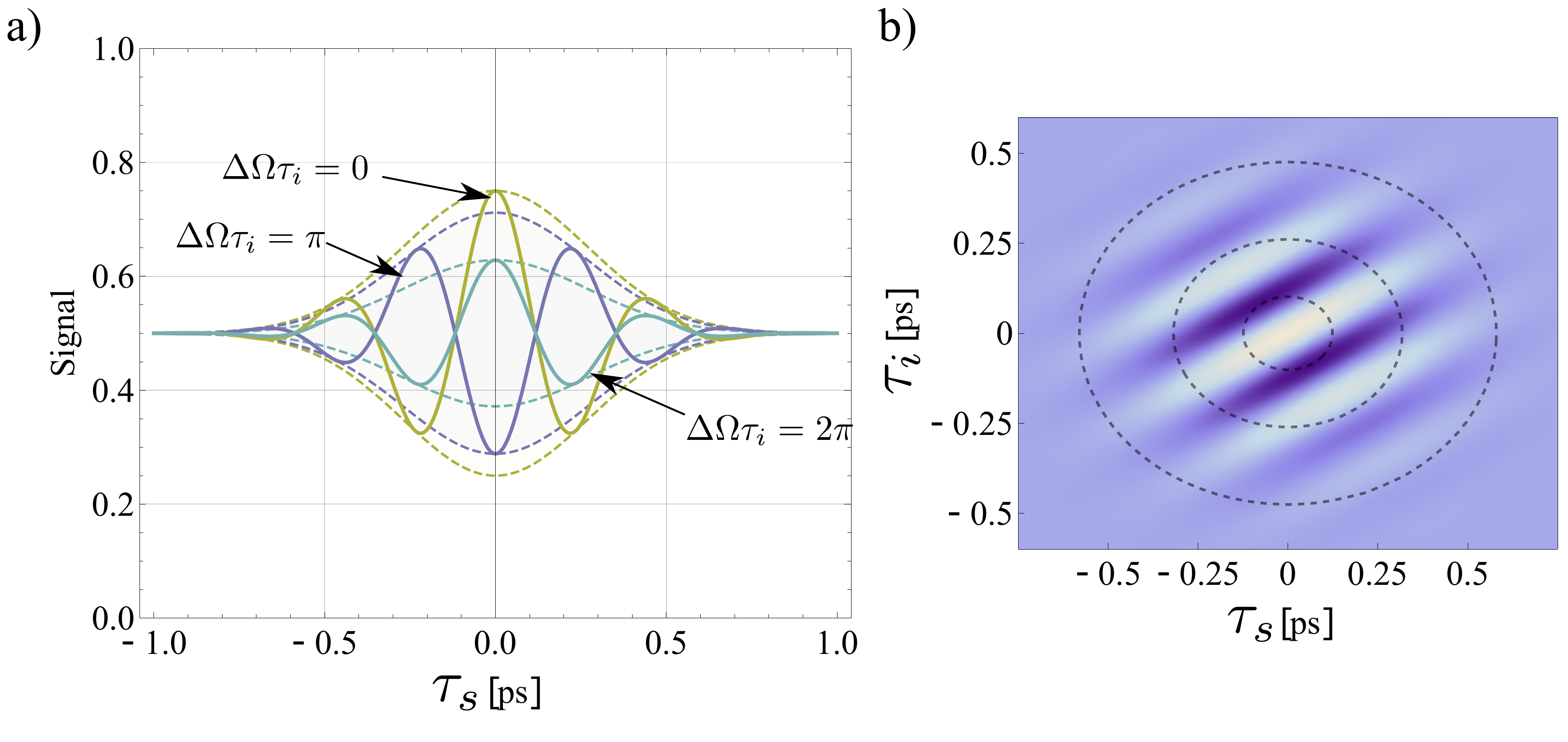}
	\caption{a) Simulated coincidence probability as a function of delay in the signal arm, using experimental parameters and for different delays on the heralding (idler) arm. The delays corresponding to phase offset $\pi$ and $2\pi$ are respectively 130 and 260 fs. b) Heatmap of the interferogram of a) with respect to both signal and idler delays. The contours correspond to 90\%, 50\% and 10\% of the envelope intensity.}\label{fig:fringes}
\end{figure}

By considering that the distributions have similar center wavelength $(\omega_{R,B},\omega_{b,r})$ and relative delays $\tau_s,\tau_i$, respectively for the signal and idler, the coincidence probability is simply given by:

\begin{align}
P_{cc}(\tau_s,\tau_i) = \frac{1}{2}+\frac{1}{4} \mathcal{E}_s(\tau_s)\mathcal{E}_i(\tau_i) \cos[\Delta\Omega\ \tau_i - \Delta\omega\ \tau_s] \label{eq:fringesfull}
\end{align}
where $\mathcal{E}_{s,i}$ denote respectively the envelope of the cross correlation function between the two sources spectral distributions projected on the signal or idler axis and $\Delta\omega = \omega_B-\omega_R$,  $\Delta\Omega = \omega_b-\omega_r$ are the frequency difference between the distributions. Note that this model works for non filtered sources where the frequency difference is zero on both dimensions.

The coincidence probability is plotted in the $(\tau_s,\tau_i)$ plane in Fig.\ref{fig:fringes}b) for Gaussian distributions with moments that correspond to the experimental values. It shows that, for a temporal offset from zero in the idler arm, the fringes visibility and phase are affected. This is depicted clearly in Fig.\ref{fig:fringes}a) where we plotted the probability for values of the idler delay of $0, 130, 260$ fs which correspond to a phase offset in \eqref{eq:fringesfull} of $0,\pi$ and $2\pi$. The fringes visibility then drops from 50\% to 25\% because of the envelope of the cross correlation function of the idler distributions. This shows how proper delay matching between the two independent idler photons is critical. The phase offset was observed during the experimental acquisition of the coincidence probability due to the difficulty to identify the location of the HOM dip in the idler arm of the interferometer. Moreover, we can see that optical path fluctuations that occur easily during a long acquisition at room temperature will result in a loss of contrast.

Note that the simulation from Fig.\ref{fig:fringes} takes into account the background level due to two photon events from each sources but does not plot the high frequency fringes. These are incorporated in the model by adding an oscillating term $\cos[\Omega_+\ \tau_i - \omega\_+ \tau_s]$ where $\omega_+ = \omega_B+\omega_R$,  $\Omega_+ = \omega_b+\omega_r$. Since that interferometric term is oscillating at a much higher frequency, any fluctuation in optical path would average it to zero, and hence it is reasonable to neglect that term.

\end{document}